\documentclass[12pt, preprint,twoside]{aastex}
\usepackage{geometry}               
\usepackage{graphicx}
\usepackage{amssymb}
\usepackage[suffix=]{epstopdf}
\usepackage{fullpage}
\usepackage{natbib}
\usepackage{subfigure}
\usepackage{epsfig}

\renewcommand{\&}{and}

\title{Spectrum-driven Planetary Deglaciation\\Due to Increases in Stellar Luminosity}
\author{Aomawa L. Shields\altaffilmark{1,2,3},
Cecilia M. Bitz\altaffilmark{4},
Victoria S. Meadows\altaffilmark{2,3},\\
Manoj M. Joshi\altaffilmark{5},
Tyler D. Robinson\altaffilmark{6,3}\\
}

\altaffiltext{1}{\bf{Corresponding author: Aomawa Shields, Department of Astronomy, University of Washington, Box 351580, Seattle, WA 98195-1580}}
\altaffiltext{2}{Department of Astronomy and Astrobiology Program, University of Washington}
\altaffiltext{3}{NASA Astrobiology Institute}
\altaffiltext{4}{Department of Atmospheric Sciences, University of Washington}
\altaffiltext{5}{School of Environmental Sciences, University of East Anglia}
\altaffiltext{6}{NASA Ames Research Center}

\begin{document}
\begin{abstract}
Distant planets in globally ice-covered, ``snowball", states may depend on increases in their host stars' luminosity to become hospitable for surface life. Using a General Circulation Model (GCM), we simulated the equilibrium climate response of a planet to a range of instellations from an F-, G-, or M-dwarf star. The range of instellation that permits both complete ice cover and at least partially ice-free climate states is a measure of the climate hysteresis that a planet can exhibit. An ice-covered planet with high climate hysteresis would show a higher resistance to the initial loss of surface ice coverage with increases in instellation, and abrupt, extreme ice loss once deglaciation begins. Our simulations indicate that the climate hysteresis depends sensitively on the host star spectral energy distribution. Under fixed CO$_{2}$ conditions, a planet orbiting an M-dwarf star exhibits a smaller climate hysteresis, requiring a smaller instellation to initiate deglaciation than planets orbiting hotter, brighter stars. This is due to the higher absorption of near-IR radiation by ice on the surfaces and greenhouse gases and clouds in the atmosphere of an M-dwarf planet. Increases in atmospheric CO$_{2}$ further lower the climate hysteresis, as M-dwarf snowball planets exhibit a larger radiative response than G-dwarf snowball planets for the same increase in CO$_{2}$. For a smaller hysteresis, planets near the outer edge of the habitable zone will thaw earlier in their evolutionary history, and will experience a less abrupt transition out of global ice cover.\end{abstract}
\maketitle
\pagebreak

\section{Introduction}
Planets orbiting beyond their host stars' habitable zones may exist in stable, globally ice-covered states \citep{Budyko1969, Sellers1969}, analogous to ``Snowball Earth'' episodes \citep{Kirschvink1992}. Exit out of Snowball Earth is often attributed to the build-up of CO$_{2}$ in the atmosphere as a result of volcanic outgassing and decreased silicate weathering \citep{Pierrehumbert2011b}. Given that the carbonate-silicate cycle on Earth is sensitive to plate tectonic speeds and mantle degassing rates \citep{Driscoll2013}, the efficiency of a similar mechanism on other planets may be variable. Without a continuously-operating carbon cycle to stabilize the climate, planets may rely on the steady brightening of their host stars over time \citep{Iben1967, Gough1981} to become hospitable for surface life. Such a scenario has been referred to as a ``cold start'' \citep{Kasting1993}. 

In earlier work we explored the effect of host star spectral energy distribution (SED) and ice-albedo feedback on the susceptibility of orbiting planets to enter snowball states \citep{Joshi2012, Shields2013}. We found that the large fraction of near-infrared (near-IR) radiation received by M-dwarf planets renders them less susceptible to snowball episodes than their counterparts orbiting stars with higher visible and near-ultraviolet (near-UV) output. This is due to the lower albedo of ice and snow at near-IR wavelengths \citep{Dunkle1956}, in addition to near-IR absorption by atmospheric CO$_{2}$, water vapor and clouds \citep{Shields2013}. Beginning with habitable surface temperatures and low planetary ice cover, a ``warm start", much larger decreases in stellar insolation (hereafter ``instellation") were required to trigger snowball conditions on M-dwarf planets than G- or F-dwarf planets. This lower climate sensitivity, as indicated by smaller, more gradual increases in ice extent for a given decrease in instellation, indicated a greater stability against snowball episodes on M-dwarf planets. 

Expanding on this previous work, here we explore a planet's climatic response from a cold start to \emph{increases} in instellation as a function of the SED of its host star. From the point where the instellation is just low enough for a planet to freeze into a snowball state, the amount of resistance to melting by increasing instellation is a measure of the hysteresis. In climate science, the degree of hysteresis is explored by forcing a system in two directions to determine if multiple stable states could exist for a given forcing parameter. 

A cold start is a possible path for an extrasolar planet orbiting beyond the outer edge of its host star's habitable zone, and Super-Earth-mass planets in these types of orbits have already been found by microlensing surveys \citep{Beaulieu2006, Beaulieu2010}. The amount of increased instellation required to melt a distant planet out of global ice cover as a function of host star SED has not been quantified, and we determine it here.

The range of instellations over which multiple stable equilibria are possible is representative of a planet's climate hysteresis. If a planet's hysteresis is high, its ice extent may remain at the equator despite significant increases in stellar flux. Once this resistance is finally overwhelmed, an abrupt change in ice extent may occur. Whether this would have a positive or negative influence on the development and evolution of life is under debate. Regardless, abrupt climate transitions could have important consequences for life, therefore knowing which planets are more likely to experience them, and which are likely to have more stable climates (with gradual climate transitions), will aid in target selection for follow-up missions.

This study examines how stellar SED influences climate stability, by comparing the amount of instellation required to melt out of a snowball state on planets orbiting M-, G-, and F-dwarf stars. We also compare how these planets exit a snowball with Earth's present level versus the maximum level of atmospheric CO$_{2}$ believed to exist at the end of Snowball Earth (0.1 bar, \citealp{Pierrehumbert2011b}). We discuss the implications of our results for planetary habitability.

\section{General Circulation Model}

We used version 4 of the Community Climate System Model (CCSM4), a fully-coupled, global climate model \citep{Gent2011}. We ran CCSM4 with a 50-meter deep, slab ocean (see e.g., \citealp{Bitz2012}), with the ocean heat transport set to zero, as done in experiments by Poulsen \emph{et al.} (\citeyear{Poulsen2001}) and Pierrehumbert \emph{et al.} (\citeyear{Pierrehumbert2011b}). The ocean is treated as static but fully mixed with depth. The horizontal resolution is 2$^\circ$. There is no land, hence we refer to it as an "aqua planet". The sea ice component to CCSM4 is the Los Alamos sea ice model CICE version 4 \citep{Hunke2008}. We made the ice thermodynamic only, and use the more easily manipulated sea-ice albedo parameterization from CCSM3 - with the surface albedo divided into two bands, visible ($\lambda \leqslant$ 0.7 $\mu$m) and near-IR ($\lambda >$ 0.7 $\mu$m). We used the default near-IR and visible band albedos  (0.3 and 0.67 for cold bare ice and 0.68 and 0.8 for cold dry snow, respectively). For more details, see Shields \emph{et al.} (\citeyear{Shields2013}).

We used composite SEDs derived from observations and models of the main-sequence stars HD128167 (F2V) and AD Leo\footnotemark{} (M3V, \citealp{Reid1995}, \citealp{Segura2005}), and the solar spectrum obtained from Chance and Kurucz (\citeyear{Chance2010}). 

\footnotetext[1]{http://vpl.astro.washington.edu/spectra/stellar/mstar.htm}

The Community Atmosphere Model version 4 (CAM4.0) divides the incident stellar (shortwave) radiation into twelve wavelength bands. Percentages of shortwave radiation outside of the range covered by these wavebands were folded into the shortest and longest wavebands, respectively, to include the full stellar spectrum for all stars. 

We have assumed an Earth-like present atmospheric level (PAL) of CO$_{2}$, H$_{2}$O, and O$_{2}$. The ozone profile was set to zero, as it has a negligible effect on the surface temperature of M- and G-dwarf planets \citep{Shields2013}. 

For each warm-start GCM simulation, the model was run for 37 years to equilibrate to the modern Earth climate at present solar instellation. We then ran our simulations for 40 years after that, with decreasing instellation from either the Sun, an M-dwarf, or an F-dwarf star. Our cold-start runs were started with the ice-covered (snowball) climate state found at the end of the warm-start runs, and the instellation was then increased over 40 years of simulation. We ran select simulations for an additional 50 years to ensure equilibration of the ice extent. We also ran simulations of M- and G-dwarf snowball planets with an atmospheric CO$_{2}$ concentration of 10\%. Assuming the CO$_{2}$ weathering feedback on a snowball planet still operates, we compare the planetary response to high amounts of CO$_{2}$ as a function of host star SED. 

M-dwarf planets in their host stars' habitable zones are likely to become synchronously rotating \citep{Dole1964, Kasting1993, Joshi1997, Edson2011}. However, given that our simulated planets are distant, frozen worlds, we assumed a 24-hr rotation period. We also used an Earth-like obliquity of 23$^{\circ}$ in our GCM simulations, an eccentricity of zero, and included no land, therefore both temperature and ice behavior are symmetric in the time-mean about the equator over an annual cycle. 

\section{Results} 
Figure 1 shows the latitudinal extent of ice and global mean surface temperature on M-, G-, and F-dwarf planets as a function of percent of the modern solar constant (1360 W/m$^2$, i.e. the present insolation on Earth), assuming both warm- and cold-start conditions. As described in Shields \emph{et al.} (\citeyear{Shields2013}), the slopes of the lines from warm-start conditions are a measure of the climate sensitivity to decreases in instellation. The shallower slope of the M-dwarf planet's ice line latitude evolution indicates a lower climate sensitivity. While the F- and G-dwarf planets become ice-covered at 98\% and 92\% of the modern solar constant respectively when started in a warm state, global ice cover does not occur on the M-dwarf planet until its instellation as been reduced to 73\% of the modern solar constant. 

Upon reaching ice-covered conditions in our warm-start simulations, we then increased the instellation on all three snowball planets in intervals of roughly 1\%. Open water first appears on the M-dwarf planet with a 9\% increase in instellation, to 82\% of the modern solar constant. The latitude of the ice line at this point is $\sim$36$^\circ$. From that point on, both initial warm- and cold-start conditions yielded similar stable ice lines, as evidenced by the overlapping values at higher instellations for the M-dwarf planet in Figure 1. Mean ice line latitudes for warm- and cold-start conditions differed in latitude by less than 0.5$^\circ$ and 0.3$^\circ$ for the M-dwarf planets receiving 85\% and 90\% of the modern solar constant, respectively.

From a cold start, the G-dwarf planet does not exhibit a non-equatorial ice line until the instellation it receives has been increased by 14\% (to 106\% of the modern solar constant) relative to the instellation that brought it into a snowball state from a warm start, assuming fixed (PAL) CO$_{2}$. At this instellation, the ice line latitude has jumped from 0$^\circ$ to 69$^\circ$. The F-dwarf planet does not exhibit non-zero ice lines until its instellation has been increased by 16\% relative to the instellation that triggered global ice cover. At 114\% of the modern solar constant, the ice line latitude has jumped from 0$^\circ$ to $\sim$84$^\circ$. Further increases in instellation yielded ice-free states for both warm- and cold-start conditions. 

Figure 2 shows northern hemisphere winter surface albedo, cloud parameters and surface temperature on all three planets in hard snowball conditions, and just prior to deglaciating. Cloud radiative forcing is the irradiance difference at the top-of-atmosphere (TOA) with and without clouds for the absorbed shortwave, outgoing longwave, or the net radiation.  In the snowball state, the shortwave cloud forcing (SWCF) is slightly positive in austral summer over most latitudes because the clouds for all three planets absorb radiation, reducing the amount of shortwave radiation reflected by surface snow and ice \citep{Cogley1984}. The SWCF is most positive on the M-dwarf snowball planet, even though it is the least cloudy among the three cases. This is due to clouds absorbing more of the near-IR radiation emitted by the M-dwarf star.

Just prior to deglaciating, the surface albedo drops significantly from $\sim$20-50$^\circ$S as the snow sublimates or melts away in summer months. Here the shortwave effect of clouds is distinctly cooling due to scattering. Because cloud particles are relatively large compared to the wavelengths of radiation from all three stars, the scattering potential does not depend on the stellar SED. Instead, the amount of cloud cover and the surface albedo are the dominant factors controlling the SWCF. The SWCF is similar among the three planets from 20-50$^\circ$S, owing to compensating effects. The M-dwarf planet is the least cloudy, however its surface is also the least reflective. Fewer clouds on the M-dwarf planet have just as sizeable of an effect as greater cloud cover on the G and F-dwarf planets.

The longwave (LWCF) and total (shortwave + longwave, TCF) cloud forcing tend to be smallest for the M-dwarf planet. This is due to the lower amount of cloud cover, which is a consequence of a squatter and weaker Hadley Cell compared to the other planets (Fig. 3, Table 1). This weaker Hadley circulation stems from a more stable atmospheric temperature profile on the M-dwarf planet (Fig. 4), and limits the transport of heat away from the tropics. This assists thawing in the tropics of the M-dwarf planet compared to the G and F-dwarf planets.

We also compared the M- and G-dwarf response to raising CO$_{2}$ to 0.1 bar - the upper limit of CO$_{2}$ expected to build up in the atmosphere as a result of decreased surface temperatures on an ice-covered planet, assuming volcanic outgassing occurs while silicate weathering is inhibited \citep{Walker1981}. This value depends on volcanic outgassing and seafloor weathering rates \citep{LeHir2008}, surface dust deposition \citep{LeHir2010}, and model parameterizations \citep{Pierrehumbert2004, Hu2011b, Abbot2012}. Regardless, the M-dwarf planet, which requires a much lower instellation to fully glaciate, exhibits a larger radiative response for the same CO$_2$ increase than the G-dwarf snowball planet receiving much larger instellation, yielding a 12\% increase in TOA absorbed shortwave flux for a 250-fold increase in CO$_{2}$ concentration, compared with a 7\% increase on the G-dwarf planet. 

\section{Discussion}

The results of our model simulations indicate that the amount of increased instellation required to melt a planet out of a snowball state is highly sensitive to host star SED. The stability and evolution of a planet's climate is likely a function of the spectral properties of its host star.

The range of instellations over which multiple distinct ice line latitudes are possible is indicative of the level of climate hysteresis on these planets. At the upper end of the range where multiple equilibria occur, the ice edge jumps from the equator to a poleward position. Above this point, a change in instellation yields comparable changes in ice edge for both warm- and cold-start initial conditions, arresting the multiplicity in stable climate states. M-dwarf planets have the smallest hysteresis and ice edge jump, G-dwarf planets exhibit intermediate ice edge jumps, and F-dwarf planets have the greatest jumps in ice line latitude and the largest climate hysteresis. 

As a result of the longer-wavelength radiation emitted by the M-dwarf star, more radiation is absorbed by surface ice and atmospheric CO$_{2}$ and water vapor, which have strong absorption bands in the near-IR. The greater shortwave heating in the atmosphere of the M-dwarf planet reduces the amount of radiation reaching the surface, providing less relative heat to drive convection and rising plumes of air parcels at the equator. This weakens Hadley circulation on the planet. While the lower Hadley circulation suppresses cloud formation and lowers the total cloud forcing on the M-dwarf planet (which may oppose deglaciation), its larger effect is to reduce heat transport from the tropics to higher latitudes. This causes temperatures to rise above freezing in the sub-tropics and surface melting to occur. The G- and F-dwarf planets, with stronger Hadley circulation and enhanced cloud formation, demonstrated a greater tendency to remain in a snowball state despite increased instellation.

In non-snowball conditions the M-dwarf planet permits a stable ice line that is $\sim$33$^\circ$ and $\sim$48$^\circ$ closer to the equator than those of the G- and F-dwarf planets, respectively. As snowball deglaciation may be highly sensitive to surface albedo \citep{Lewis2006}, and to ocean-ice albedo differences \citep{Abbot2011}, the smaller difference in ocean-ice albedo contrast likely contributes to the lower stable ice line generated on the M-dwarf planet. We ran sensitivity tests with an energy balance model with albedos and atmospheric absorption determined from a line-by-line radiative transfer model \citep{Meadows1996}, and found both the climate hysteresis and the latitude of stable ice lines to be lower for smaller ocean-ice albedo differences, and more sensitive to ice albedo differences than changes in atmospheric absorption. The inclusion of ocean heat transport, which has been shown to hasten the retreat of sea ice with increased instellation on M-dwarf planets \citep{Hu2014}, may increase the stable ice line latitudes that we have calculated here for the deglaciating M-dwarf planet.

We have assumed an eccentricity of zero and an Earth-like obliquity in this work. High-eccentricity planets receiving instellation from the Sun melt out of snowball states more easily \citep{Spiegel2010}. And planets with high obliquities \citep{Williams1997, Spiegel2009} or high-frequency obliquity oscillations (Armstrong et al. 2014, in press) are less susceptible to snowball episodes. Therefore the trend of a smaller climate hysteresis for planets orbiting cooler stars may be further amplified at high obliquity and eccentricity. 

Simulations with higher CO$_{2}$ resulted in a larger radiative response for the same increase in CO$_{2}$ on the M-dwarf snowball planet than on its G-dwarf counterpart. If a carbon cycle operates on distant planets, ice-covered M-dwarf planets would likely exhibit a greater atmospheric response to the steady build-up of volcanically-outgassed CO$_{2}$, and may require less CO$_{2}$ to deglaciate. This would lower the climate hysteresis of M-dwarf planets even further compared to planets orbiting brighter stars. Our fixed-CO$_{2}$ results therefore represent a lower limit on the difference in climate hysteresis as a function of stellar spectral type. 
 
Climate hysteresis will affect the fraction of a planet's lifetime over which it can maintain habitability. As a main-sequence (core hydrogen-burning) star ages, its luminosity increases \citep{Gough1981}. Standard solar evolution models indicate that the Sun's luminosity has increased by $\sim$30\% since its arrival on the main-sequence \citep{Newman1977, Feulner2012}. The Sun's luminosity is estimated to increase by $\sim$9\% over the next billion years \citep{Gough1981}, and the 14\% increase in instellation to melt out of global ice cover (in the absence of an active carbon cycle) would require 1.4 Gy of stellar evolution, assuming no significant changes to the atmospheric composition of the planet. This is approximately 13\% of the main-sequence lifetime of the Sun \citep{Sackmann1993}. 

M-dwarf stars, given their smaller masses (0.08-0.5 M$_\odot$), burn their fuel at much lower rates \citep{Iben1967, Tarter2007} than G- or F-dwarf stars, and so brighten more slowly. Recent models of low-mass stellar evolution predict modest luminosity increases of $\sim$0.5-1\% per billion years for a 0.4 M$_\odot$ star, depending on its age \citetext{Rushby et al. 2013; Andrew Rushby, priv.\ comm.}. Given this range and the 9\% increase in instellation required to generate open ocean on a frozen M-dwarf planet, thawing would occur in $\sim$9-18 billion years. However, this is less than 8\% at most of the main-sequence lifetime of a 0.4 M$_\odot$ star ($\sim$225 Gyr, \citealp{Guo2009}). Therefore a frozen M-dwarf planet would thaw out earlier in the evolutionary path of its host star than ice-covered planets orbiting hotter, brighter stars, providing a longer timescale for biological evolution to evolve from frozen surface conditions.

\section{Conclusions}

Using 3-D climate simulations, we have demonstrated that the climate stability and evolution of a planet depends on the spectral energy distribution of its host star. M-dwarf planets exhibit climate hysteresis over a smaller range of incident stellar radiation than planets orbiting stars with higher visible and near-UV output, as indicated by the narrower range of instellations over which multiple stable ice lines are possible. Thawing M-dwarf planets exhibit less abrupt jumps in ice line latitude, which may be more advantageous for life. An M-dwarf snowball planet is likely to melt more easily out of global ice cover as its host star ages and brightens. This is due to the combined effects of surface ice and snow absorption of the large fraction of near-IR radiation emitted by M-dwarfs, and atmospheric near-IR absorption, which weakens Hadley circulation, reducing the hysteresis of M-dwarf planets. Planets near the outer edge of the habitable zones of M-dwarf stars will become more hospitable for surface life earlier in their host stars' evolutionary paths than their ice-covered counterparts orbiting brighter stars, although this may take a longer absolute time, as an M dwarf brightens more slowly than a G dwarf. If a silicate weathering feedback operates on these cold outer worlds, increased CO$_{2}$ would further reduce the climate hysteresis on M-dwarf planets with equivalent surface temperatures as G-dwarf planets, providing increased stability against permanent low-latitude glaciation. 

\section{Acknowledgments}

This material is based upon work supported by the National Science Foundation Graduate Research Fellowship Program under Grant Nos. DGE-0718124 and DGE-1256082. This work was performed as part of the NASA Astrobiology Institute's Virtual Planetary Laboratory Lead Team, supported by the National Aeronautics and Space Administration through the NASA Astrobiology Institute under Cooperative Agreement solicitation NNH05ZDA001C. We thank Dorian Abbot and Raymond Pierrehumbert for helpful insight on this work, and an anonymous reviewer for extremely helpful comments that greatly improved the paper.

\newpage

\newpage
\linespread{1.15}

\begin{figure}[!htb]
\begin{center}
\includegraphics [scale=2.00]{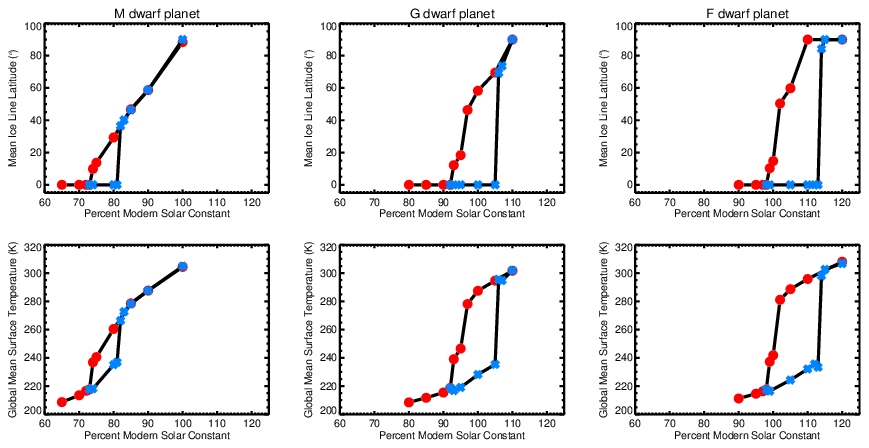}\\
\caption{Mean ice line latitude (top) and global mean surface temperature (bottom) as a function of stellar flux for an aqua planet orbiting an M-, G-, and F-dwarf star. Simulations assuming an initial warm start are in red (circles). Initial cold start simulations are in blue (asterisks).}
\label{Figure 1. }
\end{center}
\end{figure}

\linespread{1.15}

\begin{figure}[!htb]
\begin{center}
\includegraphics [scale=1.00]{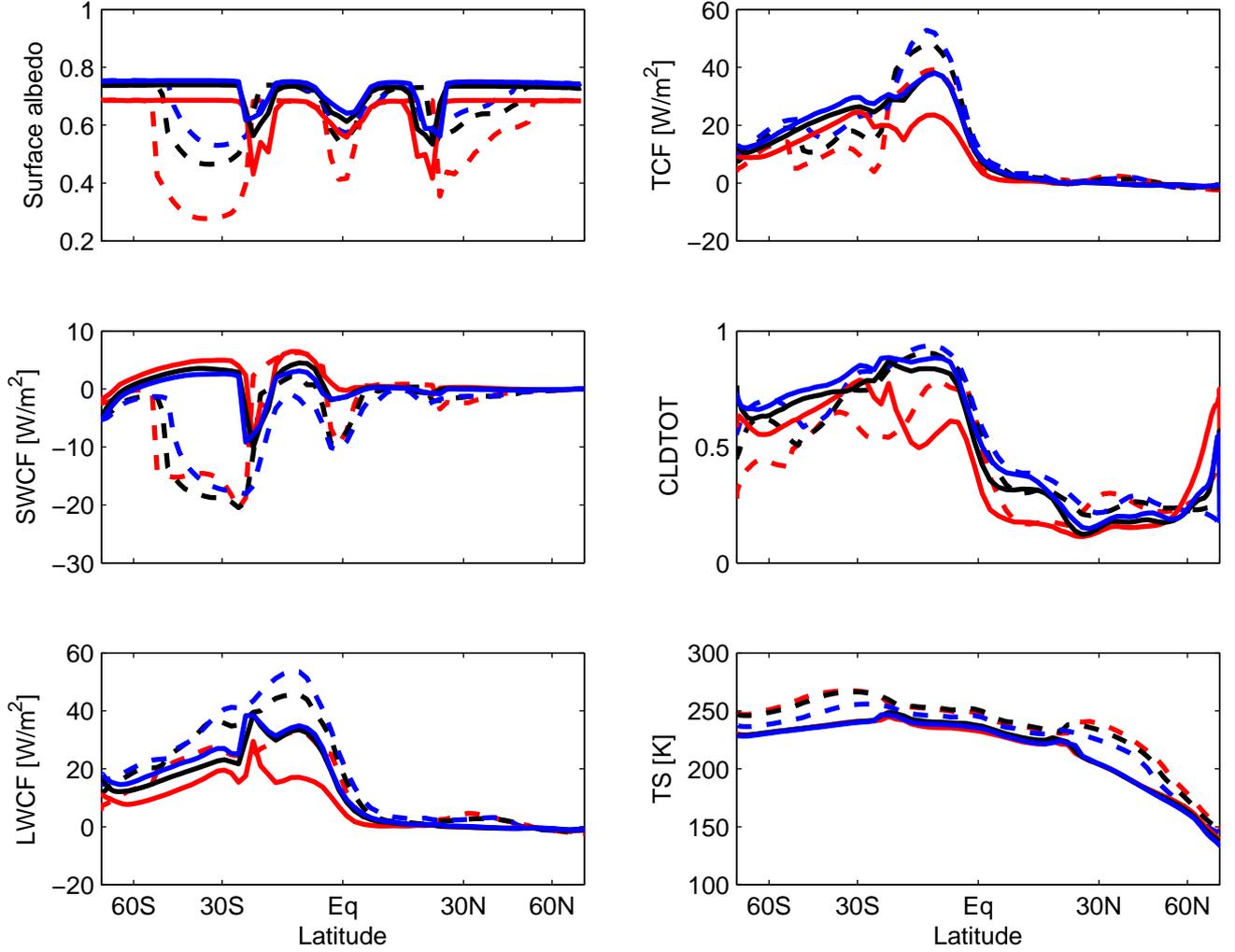}
\caption{Northern winter (December/January/February, DJF) average surface albedo, shortwave (SWCF), longwave (LWCF), and total (TCF) cloud forcing, total cloud fraction (CLDTOT), and surface temperature (TS) versus sin of latitude for M-(red), G- (black) and F- (blue) dwarf planets in a snowball climate (solid) and prior to the appearance of open ocean (dashed).}
\label{Figure 2. }
\end{center}
\end{figure}

\begin{figure}[!htb]
\begin{center}
\includegraphics [scale=1.00]{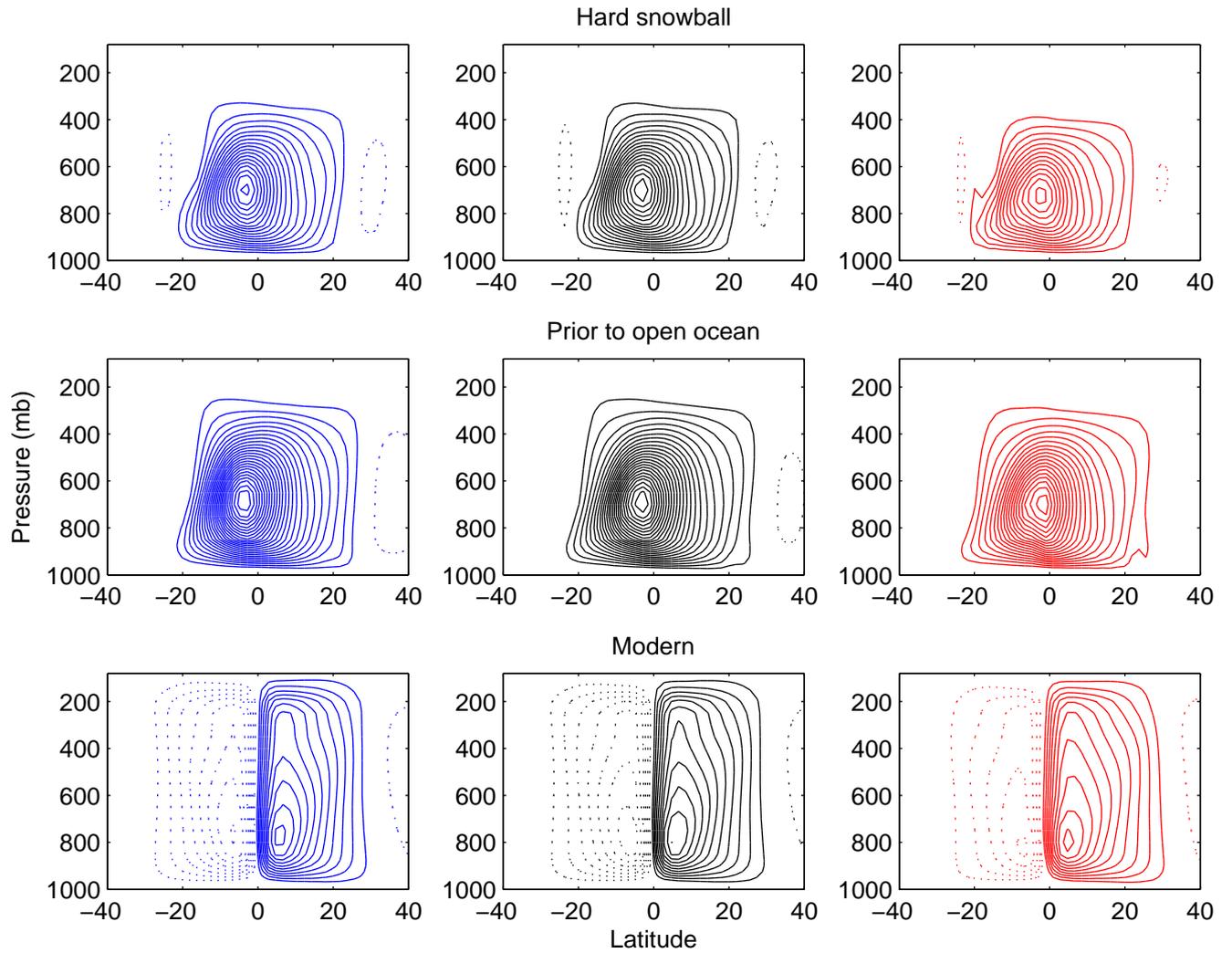}\\
\caption{DJF Meridional stream function for M- (right), G- (center), and F- (left) dwarf planets with different climates. The contour interval is 25 x $10^9$ kg/s. The zero contour interval is not shown. Dotted lines denote counterclockwise circulation.  }
\label{Figure 3. }
\end{center}
\end{figure}

\begin{figure}[!htb]
\begin{center}
\includegraphics [scale=0.40]{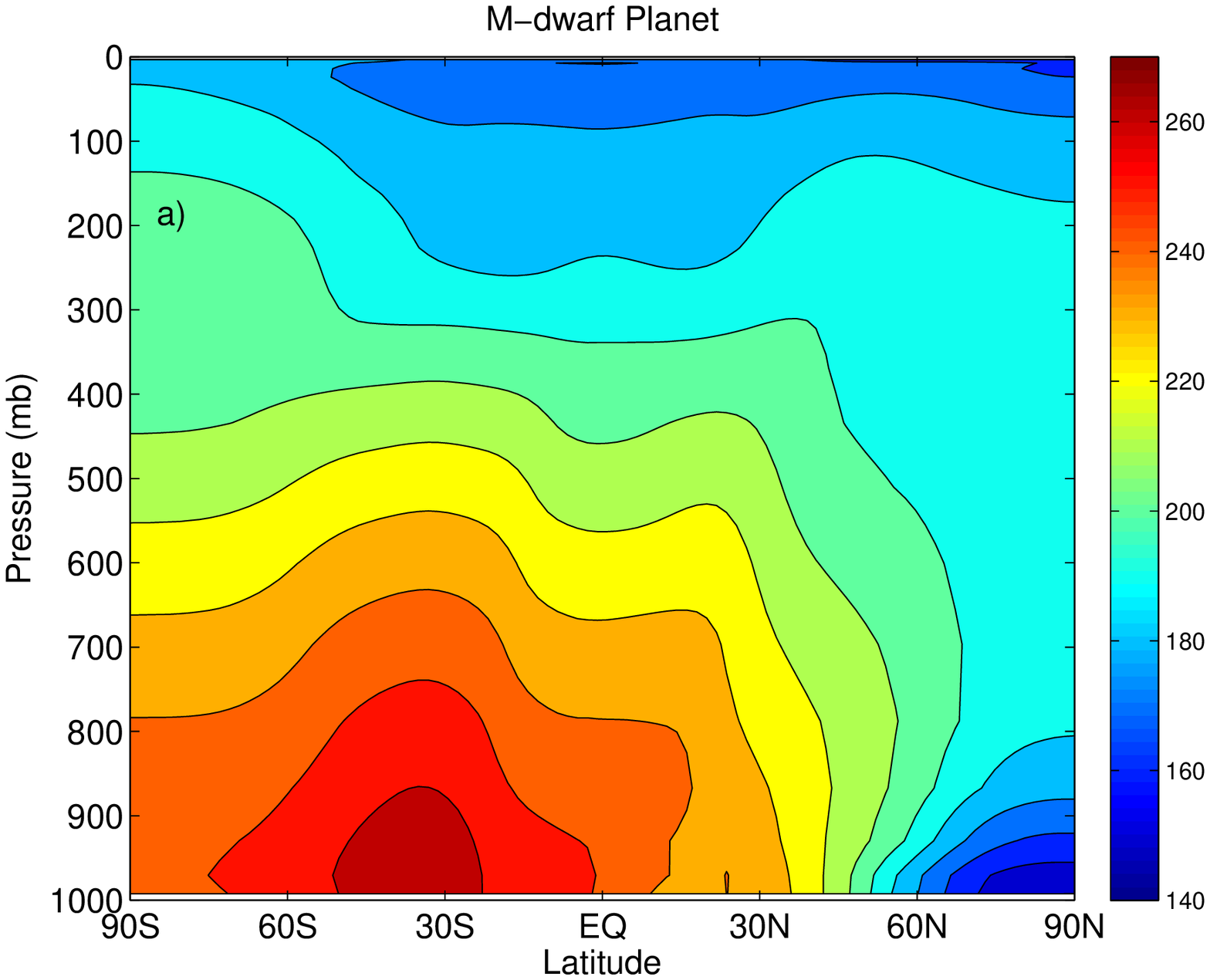}
\includegraphics [scale=0.40]{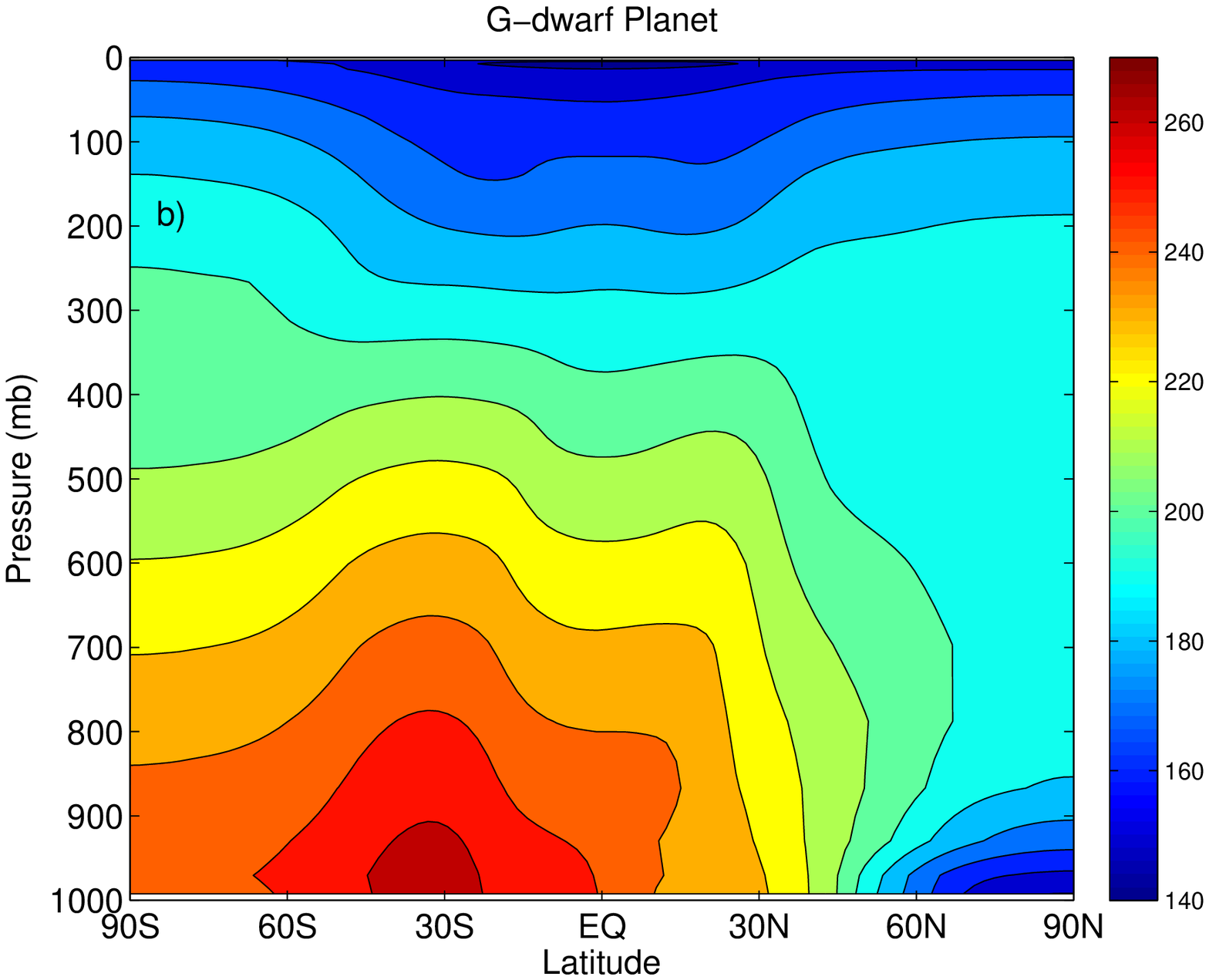}\\
\includegraphics [scale=0.40]{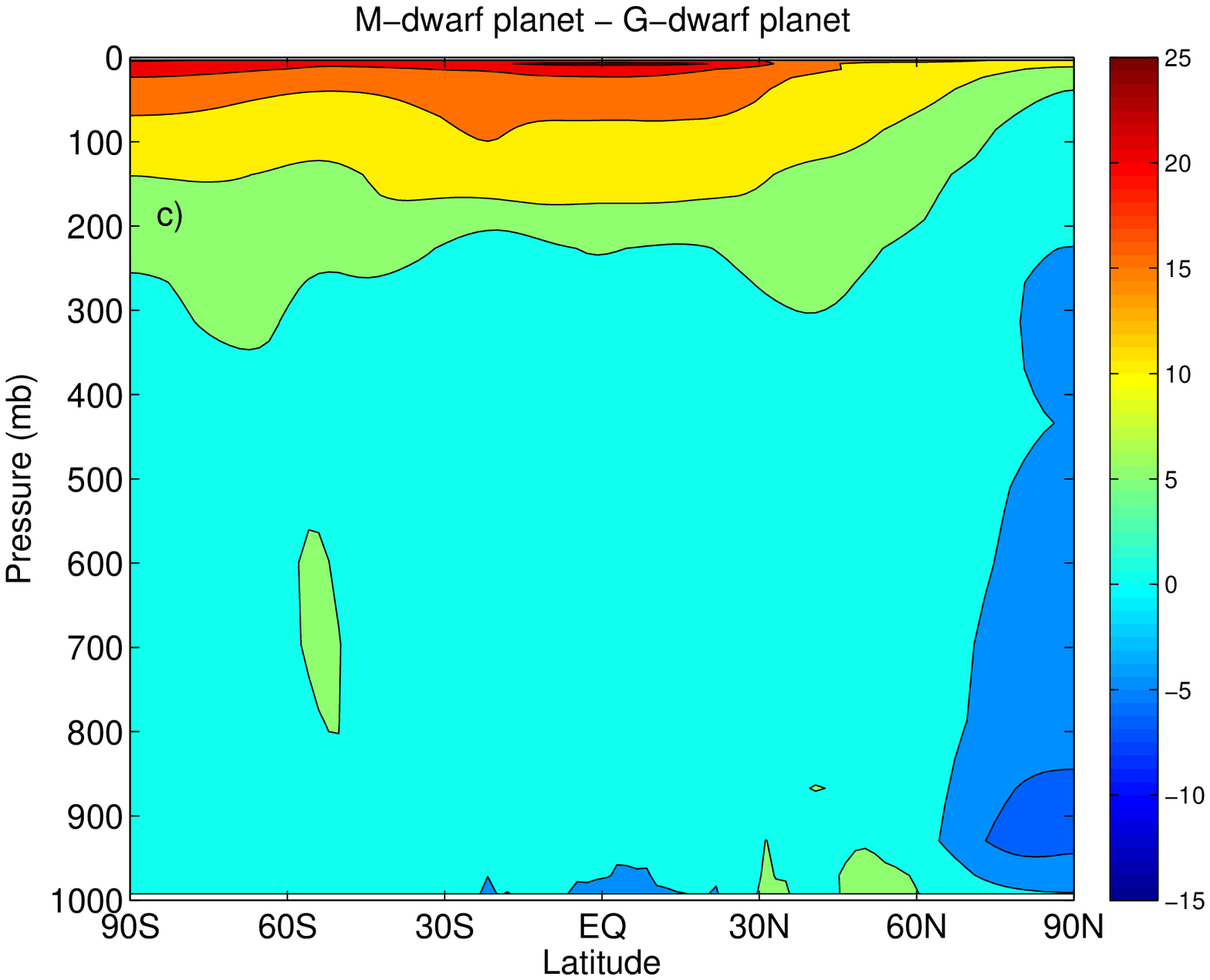}
\caption{Zonal mean DJF vertical temperature for an (a) M- and (b) G-dwarf planet prior to deglaciating; (c) Increase in vertical temperature of the M-dwarf planet, calculated by taking the difference between the M- and G-dwarf planets' atmospheric temperature profiles.  }
\label{Figure 4. }
\end{center}
\end{figure}

\begin{pagebreak}

\begin{table}[!htp] 
\caption{Boreal winter (DJF) maximum meridional stream function values for M- G-, and F-dwarf planets with different climates. Units are kg/s x 10$^9$.} 
\vspace{2 mm}
\centering \begin{tabular}{c c c c} 
\hline\hline 
Stellar Type & snowball & prior to open ocean & modern \\  [0.5ex]\\
\hline
F-dwarf & 459.24 & 644.95 & 280.74\\
\hline
G-dwarf & 469.60 & 623.22 & 275.05\\
\hline
M-dwarf & 357.29 & 489.26 & 260.50\\
[1ex]
\hline
\end{tabular} 
\label{table:nonlin} 
\end{table}
\end{pagebreak}

\end{document}